\documentclass[aps,twocolumn]{revtex4}

\usepackage{graphicx}
\usepackage{color}
\usepackage{amsmath}

\begin{document}

\title{Levitation of superconducting micro-rings for quantum magnetomechanics}
\author{Carles Navau$^{1,*}$, Stefan Minniberger$^2$, Michael Trupke$^2$, and Alvaro Sanchez$^1$}
\affiliation{$^1$ Departament de F\'{\i}sica, Universitat Aut\`onoma de Barcelona, 08193 Bellaterra, Barcelona, Catalonia, Spain}
\affiliation{$^2$  
Vienna Center for Quantum Science and Technology, Universität Wien, Boltzmanngasse 5, 1090 Vienna, Austria
}

\begin{abstract}

Levitation of superconductors is becoming an important building block in quantum technologies, particularly in the rising field of magnetomechanics. In most of the theoretical proposals and experiments, solid geometries such as spheres are considered for the levitator. Here we demonstrate that replacing them by superconducting rings brings two important advantages: Firstly, the forces acting on the ring remain comparable to those expected for solid objects, while the mass of the superconductor is greatly reduced. In turn, this reduction increases the achievable trap frequency. Secondly, the flux trapped in the ring by in-field cooling yields an additional degree of control for the system. We construct a general theoretical framework with which we obtain analytical formulations for a superconducting ring levitating in an anti-Helmholtz quadrupole field and a dipole field, for both zero-field and in-field cooling. The positions and the trapping frequencies of the levitated rings are analytically found as a function of the parameters of the system and the field applied during the cooling process. Unlike what is commonly observed in bulk superconductors, lateral and rotational stability are not granted for this idealized geometry. We therefore discuss the requirements for simple superconducting structures to achieve stability in all degrees of freedom.

\end{abstract}

\maketitle

\section{Introduction}
\label{sec.intro}

Levitation of superconductors (SCs) has been an active topic of research in the last decades due to its significant scientific and technological potential. Until recently, most of the research has been carried out for large-scale systems and in areas like transportation and energy storage \cite{Brandt1989,MoonBook,Ma2003,Hull2000,Wang2002}. A typical levitation system consists of a source of the magnetic field (e.g. one or several permanent magnets) and one (or several) superconducting objects. The performance and optimization of such devices are based mainly on the choice and control of the magnetic field sources, on the optimization of the superconducting properties and on the election of an advantageous geometry for the superconductor. Typically, a set of bulk high-$T_c$ superconductors are used as levitators, each of them with sizes on the order of a few centimetres \cite{Navau2013}.  As a rule of thumb, levitation forces are enhanced by increasing the critical-current density of the SCs and the gradients of external fields; stability is increased by an adequate tuning of the field distribution and the cooling procedure \cite{delvalle2007}. At these scales, apart from the geometry, the key parameter of the superconductor is the critical-current density $J_c$, which is related to the internal vortex structure as a macroscopic averaged quantity. The critical-state model is commonly used to describe the superconductor response \cite{Navau2013}. The calculation of the current-density distribution inside the superconducting materials is the basic step from which one can evaluate all the relevant parameters such as force, stability, or energy.

Recently, a new application of levitating superconductors has emerged in the context of quantum magnetomechanics, where magnetically trapped superconducting objects can create low-loss mechanical oscillators with long coherence times \cite{Romero-Isart2012,Cirio2012,Johnsson2016,Pino_2018,Vinante2019}. A typical system consists of a  superconducting object levitated in a specifically designed magnetic field, which creates a magnetic trap for the superconductor. The objective of these proposed experiments is to perform ground-state cooling of these objects, allowing for the generation of non-classical center-of-mass motional states. In these systems, the superconductor sizes are expected to be on the order of a few tens of micrometers or less. At these scales, the basic principles of superconductivity still hold, but some assumptions generally taken for granted macroscopically should be addressed in more detail. In particular, macroscopic approximations, such as the critical-state model, are not applicable and other approaches need to be studied.

In magnetomechanical experiments, the cooling process of the center-of-mass motion typically starts when the SC is already stabilized in the magnetic trap. Thus, the first step is to levitate a superconductor in a given configuration of the external magnetic field. The presence of external magnetic field during the cooling below $T_c$ (field cooling process, FC) can result in a substantially different current distribution in the SC with respect to the case where the superconductor is cooled in the absence of an external magnetic field (zero-field cooling process, ZFC), yielding different levitation forces and stability. Also, the displacement of the superconductor from the cooling position to the final trapping position can modify the current distribution and, thus, the force. At macroscopic scales, the effects of cooling procedures have previously been considered in several works \cite{Navau2004,Dias2010}. 

An important property of superconductors with non simply connected geometries, like rings, is that the flux that threads a hole surrounded by a superconducting region has to be conserved. At the macroscopic scale, one needs to evaluate the current distribution inside the rings; the flux conservation through the hole is a consequence of the resulting flux distribution \cite{Navau2005,Bartolome2005}.

At the microscopic scale, the exact current distribution inside the ring can become irrelevant and the flux conservation can be used as the condition for evaluating the total circulating current. The latter becomes then the relevant quantity.

While the microscopic details of the current distribution throughout the superconducting material  are key for understanding the lateral and angular stability, this circulating current becomes the relevant quantity for the motion along the axis of the ring. In most experiments, this is the direction of interest.

In this work, we develop a comprehensive analytical theory for the levitation of superconducting micro-rings and demonstrate how their use can be very advantageous for applications in quantum magnetomechanics. One advantage for using rings in magnetomechanical experiments, with respect to single-connected geometries such as spheres \cite{Romero-Isart2012,Hofer_2019}, arises from the fact that a ring with the same radius as the sphere would have much less weight. However, the current induced to expel the field from the levitator is not greatly reduced, resulting in a similar force.

We study different scenarios for cooling and trapping small superconducting rings in the presence of external fields. The flux conservation equation is used to derive analytical expressions for the circulating currents and to evaluate the interaction energy between the magnetic field source and the levitating superconductor. We focus on two cases of interest. The first, a magnetic landscape created by a pair of anti-Helmholtz coils, is introduced in Section III. This has practical importance since it has been proposed for actual quantum magnetomechanics experiments \cite{Romero-Isart2012}. 
The second studied case is a superconducting ring coaxial with the field created by a point dipole.
We find that for a given cooling position, a single parameter is enough to characterize the dynamics of the system along the rings' axis in this case. We derive the equilibrium positions and study its vertical stability for the ZFC (section IV) and FC (section V) cases.
We discuss more qualitatively the lateral and angular stability, as well as other possible configurations in section VI. 
Conclusions are presented in Section VII. 

\section{Model for the superconductor ring}
Consider a superconducting ring of mean radius $R$ through which a total current $I$ is circulating, so that the cross-section is a circle of radius $a\ll R$. The self-inductance of the ring can be regarded as \cite{LandauBookElectrodynamics}

\begin{equation}
L=\mu_0 R \left[\log\left(\frac{8R}{a}\right)-b\right],
\label{eq:selfL}
\end{equation}
where $\mu_0$ is the vacuum permeability, and $b$ is a dimensionless constant that accounts for the particular current distribution across the cross-section of the superconductor: if the current is uniformly distributed over the cross-section, $b\simeq 1.75 $, and if it flows over the surface, $b\simeq 2$. Since $a\ll R$, the particular distribution of current inside the ring hardly affects the self-inductance. Thus, the total circulating current $I$ is the relevant parameter. Note that the inductance $L$ depends on the radius of the ring as $L\sim R \log R$. In this paper, we assume that the superconducting ring is in the Meissner state with a London penetration depth $\lambda$ smaller than all other relevant dimensions. This allows us to set $b=2$. 

Whilst the ring is in the superconducting state, the magnetic flux threading it is maintained \cite{TinkhamBook}. We shall distinguish between the ZFC regime, in which the ring has reached the superconducting state far from any field source, so that the flux that threads it is zero and will be zero after any subsequent displacement; and the FC regime, where the SC has been cooled when some flux was threading it, a  flux that will remain after movements. Except otherwise indicated, the superconducting ring has the axis coincident with the $z$-axis.  The levitated superconductor has a total mass of $M$. Gravity has direction $-\hat{\bf z}$. We use standard cylindrical coordinates $(\rho,\phi,z)$.

In the limit $\lambda \ll a$ assumed throughout this work, the flux threading the superconducting ring is quantized in multiples of the flux quantum $\Phi_0$. If one is dealing with smaller dimensions such that $\lambda$ is no longer negligible, it is the fluxoid (which includes not only the flux threading the ring but also a contribution of the currents in the superconductor volume) the quantized magnitude \cite{ClemBrandt2004}.

\section{A small ring in the field of an anti-Helmholtz coil}

The first case we analyze corresponds to a superconducting ring levitating in the field created by a pair of anti-Helmholtz coils (AHC). Such field configuration, using a levitated superconducting sphere, has been recently proposed for improving the performance in quantum systems with respect to optical counterparts (e. g. longer coherence times), enabling in this way access to a completely new parameter regime of macroscopic quantum physics \cite{Romero-Isart2012,Hofer_2019}.

We consider AHC coils with radii $C$ (and, thus separation between parallel coils $C$), coaxial with the $z$-axis and centered at the origin, and with a circulating current $I_0$. The magnetic field created by the AHC, ${\bf B}_{AHC}$ and the vector potential, ${\bf A}_{AHC}$ at the central region ($\rho,|z|\ll C$) can be approximated by 

\begin{eqnarray}
{\bf B}_{AHC}(\rho,z)&=& \mu_0 \frac{24}{25\sqrt{5}}\frac{I_0}{C^2} (-\rho \hat{\boldsymbol \rho}+2z \hat{\bf z}), \label{eq:Bahc}\\
{\bf A}_{AHC}(\rho,z)&=& \mu_0 \frac{24}{25\sqrt{5}}\frac{I_0}{C^2} z \rho \hat{\boldsymbol \phi}. \label{eq:Aahc}
\end{eqnarray}

We begin our discussion by assuming that the superconducting ring of radius $R\ll C$ is located at the origin, in the absence of any applied field or current. The flux threading the SC ring is zero. We also consider that the ring can be moved in the vertical direction, maintaining the condition $|z|\ll C$. The current induced in the ring can be evaluated from the flux-conservation equation as

\begin{equation}
2\pi R A_\phi(R,z) + L I = 0.
\label{flux}
\end{equation}

Here $A_\phi(R,z)$ is the $\phi$ component of the magnetic vector potential evaluated at the position $(R, z)$. The first term in Eq. \eqref{flux} is the flux threading the superconducting ring due to the field produced by the AHC coils and the second one the flux due to the current circulating in the ring. 

From Eqs. \eqref{eq:Aahc}-\eqref{flux}, the current in the ring is
\begin{equation}
I =-\Theta I_0 \frac{z}{R}, 
\end{equation}
where we have defined the dimensionless constant
\begin{equation} \Theta\equiv\mu_0\frac{48\pi}{25\sqrt{5}}\frac{R^3}{C^2\, L},
\label{eq:theta}
\end{equation}
that geometrically characterizes the system.

The $z$-component of the Lorentz force, $F_z$, acting on a ring centred on the axis of the field is
\begin{equation}
 F_z=-\Theta^2 I_0^2 \frac{L}{R^2} z \, \hat{{\bf z}},
\end{equation}
yielding a spring constant for vertical displacement
\begin{equation}
\kappa_z \equiv -\frac{\partial F_z}{\partial z}= \Theta^2 I_0^2 \frac{L}{R^2}.
\end{equation}

$\kappa_z$ is always positive, which indicates a vertical stable levitation. Both the force and the spring constant depend on $R$ as $\sim R^3 (\log R)^{-1}$, and on the AHC radius as $\sim C^{-4}$. The dependencies for a solid sphere are $\sim R^3$ and $\sim C^{-4}$, respectively \cite{Romero-Isart2012}, showing no significant decrease of the force value despite the large reduction in weight for a ring.

The vertical frequency of this trap is obtained from the stiffness coefficient as

\begin{equation}
\omega_z =  \frac{I_0}{R} \Theta \sqrt{\frac{L}{M}}.
\end{equation}

Following Eqs. \eqref{eq:selfL} and \eqref{eq:theta}, $\omega_z$ depends on $R$ as $\omega_z \sim R^{3/2}(\log R)^{-1/2}$.

The vertically stable position of the levitation force can be found either by equating the magnetic force to the gravity or by minimizing the energy of the system $E$ with respect to the vertical position $z$. $E$  can be described from the interaction of the field of the AHC with an induced dipole of magnetic moment $I\pi R^2 \hat{\bf z}$. Setting zero potential gravity at $z=0$ we have

\begin{equation}
E=M g z - \frac{1}{2} I \pi R^2 \hat{{\bf z}} \cdot {\bf B}_{AHC}(0,z),
\end{equation}
where $g$ is the gravity constant. The equilibrium position is found to be
\begin{equation}
z_{eq}=-\frac{M \, g\, R^2}{I_0^2 \Theta^2 L}.
\end{equation}

If the ring is cooled below the critical temperature at a vertical position $z_{FC}\neq 0$ where some flux is threading it before cooling (FC case), this flux is maintained after subsequent movements. The flux conservation equation for this case is

\begin{equation}
2\pi R A_\phi(R,z) + L I = 2\pi R A_\phi(R,z_{FC}).
\end{equation}

Owing to the linear dependence of the potential vector on the vertical position (within the approximations considered here), the above expressions for the ZFC case remain valid with the change $z\rightarrow z-z_{FC}$ or  $z_{eq} \rightarrow z_{eq}-z_{FC}$. Actually, this result permits selecting the desired value for the levitation position $z_{eq}$ by adjusting the cooling position $z_{FC}$, in an actual experimental setting.

\section{Dipole and zero-field cooled superconductor}
\label{sec.dipole+zfc}

We now apply the theoretical framework to the case of a ring levitating in the field of a magnetic dipole. We start here with the ZFC case and treat the FC case in the next section.

Consider a point dipole ${\bf m}=m {\hat {\bf z}}$ located at the origin of coordinates and the superconducting ring axially symmetric with the $z$-axis, its center located at a position $d {\hat {\bf z}}$. In the ZFC case, the cooling distance is very large, so that the ring becomes superconducting in the absence of any applied field and the flux initially threading the SC ring is zero. 
This zero flux value is maintained when the SC is vertically descended towards the point dipole. The current $I$ that flows in the superconducting ring in order to maintain this flux is found from the zero-flux threading condition, as in Eq. \eqref{flux},

\begin{equation}
2\pi R A_\phi(R,d) + L I = 0.
\label{eq.flux1}
\end{equation}

Here $A_\phi(R,d)$ is the $\phi$ component of the magnetic vector potential created by the dipole and evaluated at the position ${\bf r}=R \, \boldsymbol{\hat{\rm \rho}} + d\,\boldsymbol{\rm \hat{z}}$.

The vector potential created by the point dipole is

\begin{equation}
{\bf A}(R,d)=A_\phi(R,d) \,\boldsymbol{\hat{\phi}} =\frac{\mu_0}{4\pi}\frac{m R}{(d^2+R^2)^{3/2}} \boldsymbol{\hat{\phi}}.
\end{equation}

Thus, the current flowing in the superconducting ring will depend on the distance $d$ from the dipole as

\begin{equation}
I=-\frac{\mu_0 m R^2}{2L (d^2+R^2)^{3/2}}.
\label{eq.I}
\end{equation}

The energy of the levitated ring can be evaluated from the Lorentz force acting over the current in the ring due to the field created by the dipole or, alternatively,
from the interaction between the field created by the superconducting ring ${\bf B}_{SC}$ and the dipole. Including the gravitational potential (set as zero at the $z=0$ plane), we obtain

\begin{equation}
E= M g d + \frac{2\pi^2 R^2}{L} (\Delta A_\phi)^2 = M g d - \frac{1}{2} {\bf m}\cdot {\bf B}_{SC},
\label{eq.energy1}
\end{equation}
where $\Delta A_\phi$ represents the variation (with respect to the cooling position --infinite in the present case) of the vector potential generated by the dipole and evaluated at the ring position.  Note that the term $\frac{2\pi^2 R^2}{L} (\Delta A_\phi)^2$ equals $\frac{1}{2L} (\Delta \Phi)^2$, where $\Delta \Phi$  is the variation of the magnetic flux threading the ring due to the field created by the dipole. In the present case

\begin{equation}
E= M g d + \frac{\mu_0^2 m^2 R^4}{8L (d^2+R^2)^3}.
\label{eq.Edipole}\end{equation}

Normalizing the positions to the radius of the ring, $\zeta=d/R$, and the energy to $E_0=\mu_0^2 m^2/8LR^2$, one can describe the above system with a normalized energy $e=E/E_0$ as

\begin{equation}
e = \alpha \zeta +\frac{1}{(1+\zeta^2)^3},
\label{eq.e}
\end{equation}
where the dimensionless constant 

\begin{equation}
\alpha=\frac{8 L M g R^3}{\mu_0^2 m^2}
\end{equation} 
is a \emph{single} parameter characterizing the system. $E_0$ is the energy of the superconducting ring when moved from infinity (ZFC) to the $z=0$ position.

The energy $e$ has a minimum as a function of $\zeta$ only if 

\begin{equation}
\alpha < \frac{1029 \sqrt{7}}{2028} \equiv \alpha_c \simeq 1.329.
\label{eq.alphacrit}
\end{equation}

This inequality sets a condition for the stable levitation of ZFC rings with the considered dipole. If the radius of the ring is fixed, the mass of the ring should be less than $\simeq 0.33 m^2\mu_0^2/(g L R^3)$. If the levitating material is fixed (with a given density), then the above condition sets a maximum radius for the levitated ring.

For a given $\alpha<\alpha_c$, the normalized position of the stable levitation point, $\zeta_{eq}$ can be found as the largest of the two solutions of the equation

\begin{equation}
\alpha=\frac{6\zeta_{eq}}{(1+\zeta_{eq}^2)^4},
\label{eq.alpha}
\end{equation} 
which has to be solved numerically. The vertical oscillation spring constant in this potential well can be obtained from the second derivative of the energy, evaluated at the equilibrium position, $\kappa_z=(1/2)\frac{\partial^2E}{\partial d^2}\big|_{d=R\zeta_{eq}}$. With Eq. \eqref{eq.Edipole}, the oscillation frequency can be expressed as

\begin{equation}
\omega=\omega_0 \sqrt{\frac{3(7\zeta_{eq}^2-1)}{(1+\zeta_{eq}^2)^5}},
\end{equation}
being $\omega_0=\sqrt{E_0/2R^2M}=\frac{\mu_0 m}{4R^2} \sqrt{\frac{1}{M\,L}}$ ($\zeta_{eq}$ is the largest solution of Eq.\eqref{eq.alpha}).

\begin{figure}
	\centering
		\includegraphics[width=0.45\textwidth]{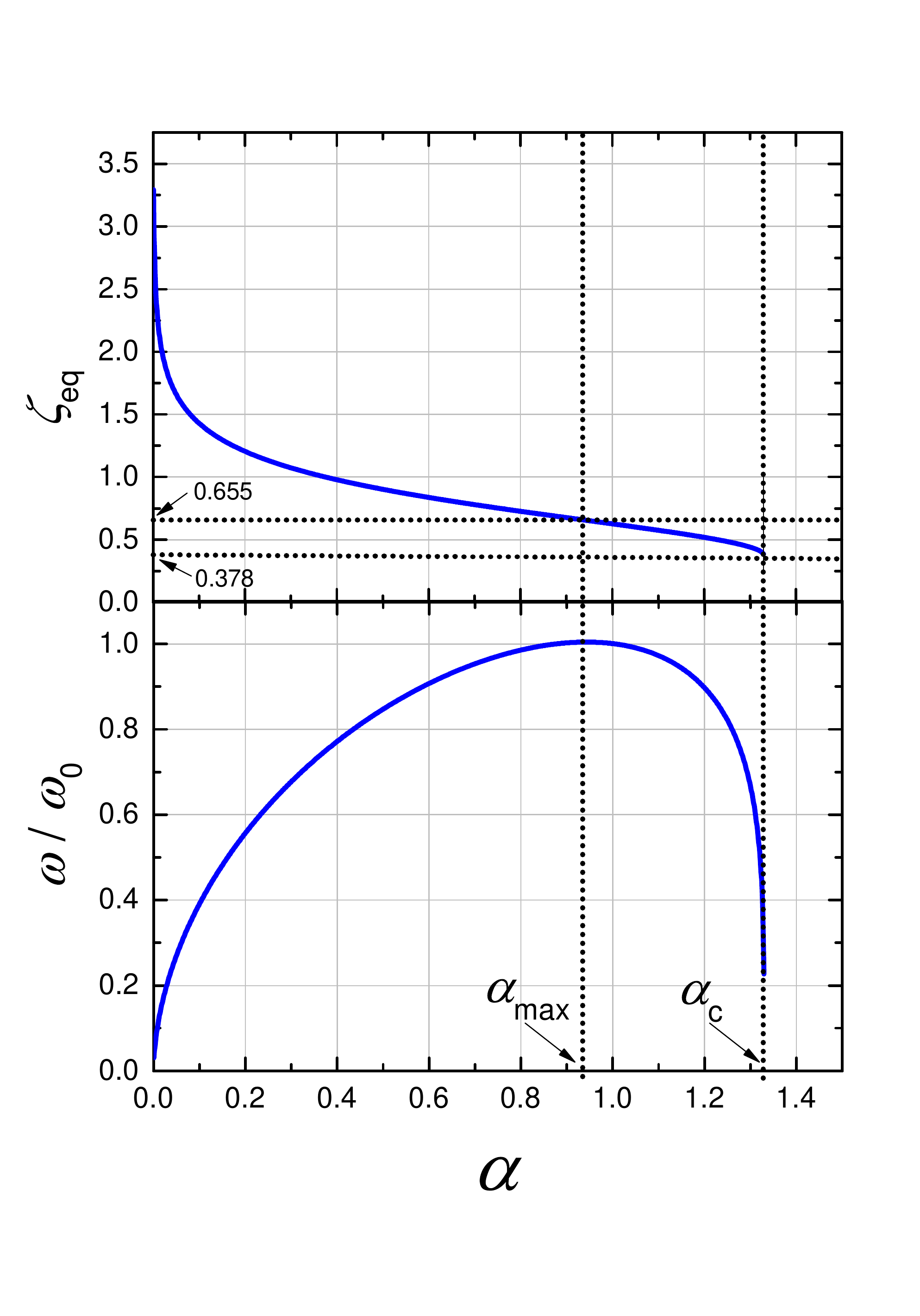}
	\caption{(a) Equilibrium position and (b) frequency of the potential well for a superconducting ring levitating in the field of a magnetic dipole, as a function of the $\alpha$ parameter characterizing the levitation (see text).  Normalization values are defined in the text.}
	\label{fig.zfc}
\end{figure}

In Fig. \ref{fig.zfc} we plot the equilibrium positions $\zeta_{eq}$ and the frequency of the trap $\omega$, as a function of $\alpha$. When $\alpha\rightarrow 0$, the equilibrium position tends to infinity as $\sim \alpha^{-1/7}$. $\alpha\rightarrow 0$ could be interpreted as a massless object or as a very large-radius ring. In addition, the frequency tends to zero as $\sim \alpha^{-4/7}$. When $\alpha\rightarrow\alpha_c$ the equilibrium position tends to $\zeta_{eq}\rightarrow 1/\sqrt{7}\simeq 0.378$ and the frequency of the trap goes to zero, corresponding to the limit in which the levitation becomes unstable. Interestingly, between these two limits there is a maximum in the stability of the force as a function of the parameter $\alpha$. This maximum is achieved when the equilibrium position is $z_{eq}=\sqrt{3/7}\simeq 0.655$, corresponding to $\alpha_{\rm max}\simeq 0.935$ (after Fig. \ref{fig.zfc}). The maximum value for the trapping frequency is $\omega_{max}=\omega_0 (49\sqrt{21})/(50\sqrt{20}) \simeq 1.004 \omega_0$. 

The fact that there exists an optimum frequency (maximum stability) means that, for a given superconducting material (a given density), there is an optimum radius that can be found from the definition of $\alpha$. Alternatively, if the geometry is also fixed, for a mass less than that needed for achieving the optimum $\alpha$, it could be interesting in practice to ballast the ring for increasing the stability.

\section{Dipole and field-cooled superconductor}
\label{sec.dipole+fc}

We consider now that the superconducting ring has been cooled below $T_c$ at a given position $d_{FC} {\bf \hat{z}}$. Now there is a flux threading the ring $\Phi_{FC}=2\pi R A_\phi(d_{FC})$ that will be maintained after subsequent movement of the SC ring when already cooled. The flux conservation equation becomes

\begin{equation}
2\pi R A_\phi(d) + L I = \Phi_{FC},
\label{eq.flux2}
\end{equation}
from which the current circulating in the ring is (defining $\zeta_{FC}=d_{FC}/R$ and $I_0 = \mu_0 m / 2 L R$)

\begin{equation}
I = I_0 \left(\frac{1}{(1+\zeta_{FC}^2)^{3/2}}-\frac{1}{(1+\zeta^2)^{3/2}}\right).
\label{eq.Ifc}
\end{equation}

Eq. (\ref{eq.energy1}) also holds in this case. The energy of the levitated ring can now be written in terms of the $\alpha$ parameter and on the $\zeta_{FC}$ value. It is found that

\begin{equation}
e=\alpha \zeta + \frac{1}{(1+\zeta^2)^3} - \frac{2}{(1+\zeta^2)^{3/2}(1+\zeta_{FC}^2)^{3/2}}.
\label{eq.efc}
\end{equation}

\begin{figure}
	\centering
		\includegraphics[width=0.5\textwidth]{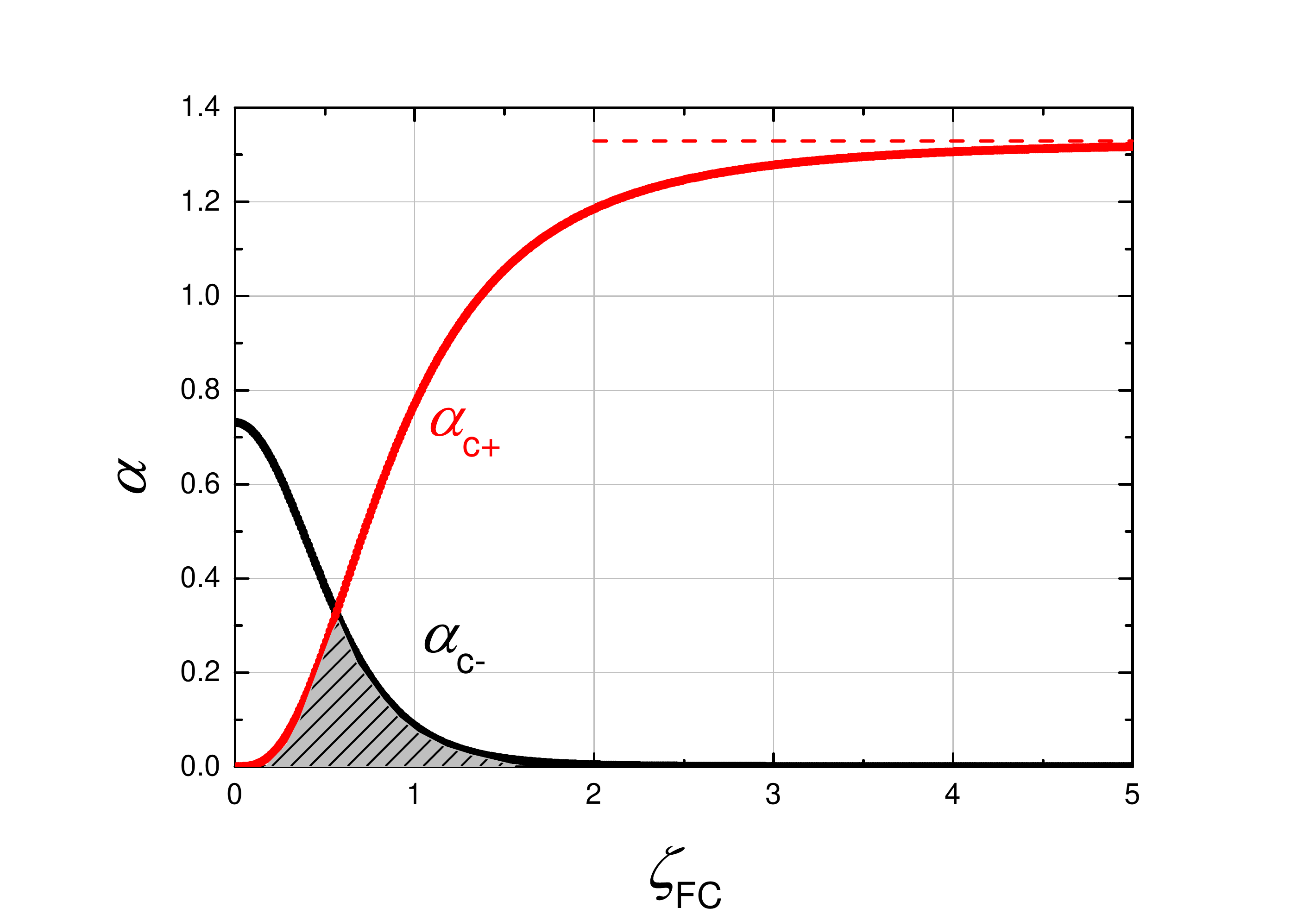}
	\caption{For a superconducting ring levitating in the field of a magnetic dipole, critical values for the $\alpha$ parameter for achieving positive equilibrium position ($\zeta>0$, red) and negative one ($\zeta<0$, black). The shadowed area corresponds the ($\alpha,\zeta_{FC}$) values that results in two equilibrium positions.}
	\label{fig.alphacrit}
\end{figure}

\begin{figure}
	\centering
		\includegraphics[width=0.5\textwidth]{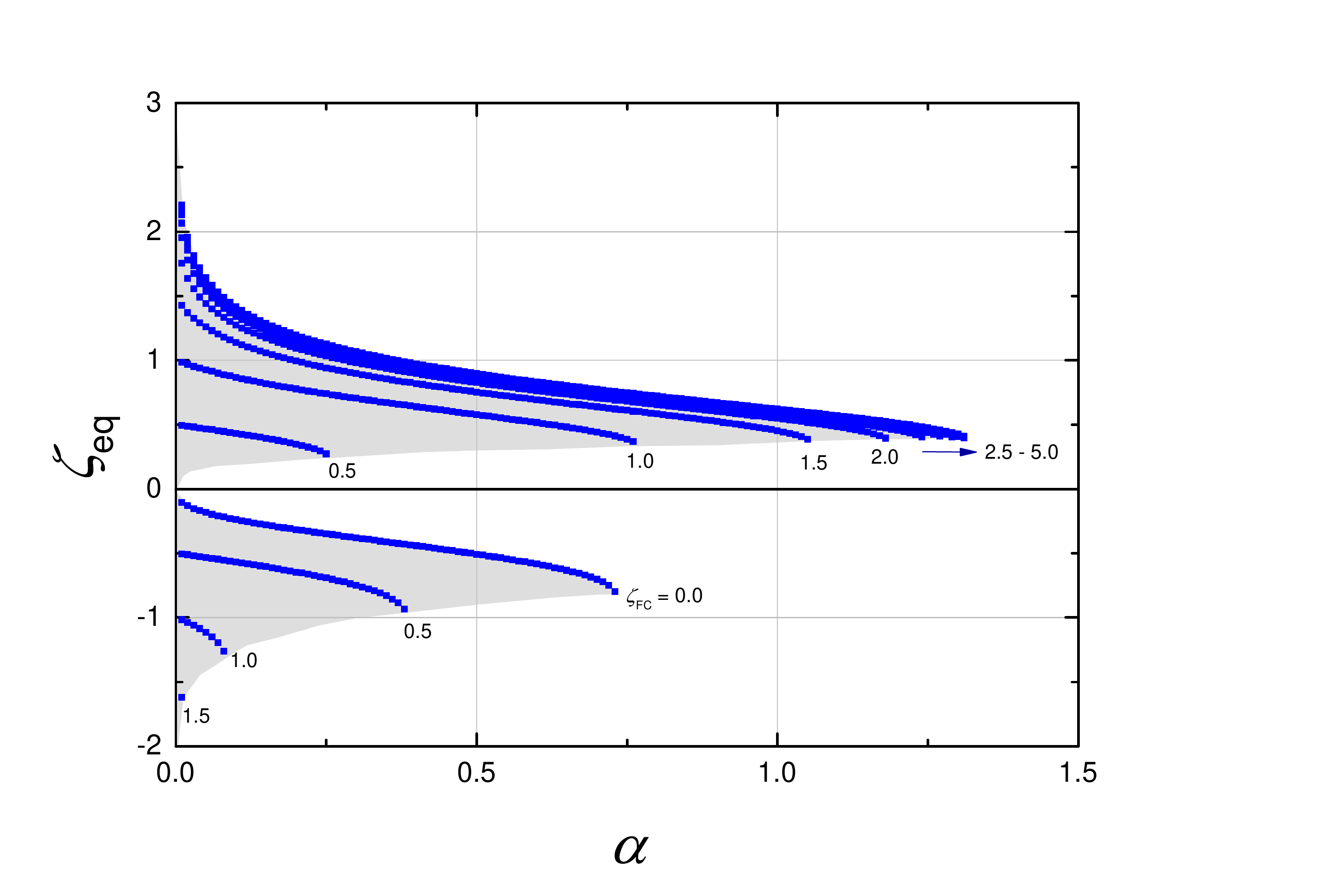}
	\caption{For a superconducting ring levitating in the field of a magnetic dipole, equilibrium positions as a function of $\alpha$ for different cooling positions (indicated in the plot). The shadowed regions correspond to all the  possible equilibrium positions for any $\zeta_{FC}$ value. Note that for some values of $\zeta_{FC}$ and $\alpha$ corresponding to the shadowed region in Fig.\ref{fig.alphacrit} there are two equilibrium positions.}
	\label{fig.equilibriumfc}
\end{figure}

Without loss of generality, we consider only positive field cooling positions ($\zeta_{FC}>0$). Now, depending on the values of $\alpha$ and $\zeta_{FC}$, the energy [Eq. \eqref{eq.efc}] can have one minimum at $\zeta<0$, one minimum at $\zeta>0$, two minima (one at $\zeta>0$ and another at $\zeta<0$), or no minima at all. Actually, for each $\zeta_{FC}$, one can define a critical $\alpha$  for having positive equilibrium positions, $\alpha_{c+}$ and another critical $\alpha$ for achieving equilibrium positions at negative $\zeta$'s, $\alpha_{c-}$. These values can be found numerically. The functions $\alpha_{c+}(\zeta
_{FC})$ and $\alpha_{c-}(\zeta_{FC})$ are plotted in Fig. \ref{fig.alphacrit}. We observe that for $\zeta_{FC}\rightarrow\infty$ we recover the ZFC case where only equilibrium at positive $\zeta$'s is possible when $\alpha<\alpha_{c+}(\infty) = \alpha_c \simeq 1.329$. $\alpha_{c-}\rightarrow 0$ in this limit. However, when the cooling position is zero (that is, when the ring is cooled with the dipole in its centre), there can be only one equilibrium position at negative $\zeta$'s. This is because the current in the ring creates an attractive force to the dipole after any movement. The equilibrium can be achieved only when the movement is made to negative positions, where the attraction to the dipole opposes gravity. Interestingly, for some values of $\zeta_{FC}$ and $\alpha$ (shadowed region in Fig. \ref{fig.alphacrit}) one can have two equilibrium positions, one at positive and the other at negative $\zeta$'s. In Fig. \ref{fig.equilibriumfc} we show the numerically evaluated equilibrium positions as a function of the $\alpha$ parameter, for different cooling distances $\zeta_{FC}$. The shadowed area in Fig. \ref{fig.equilibriumfc} includes all possible equilibrium positions for all possible $\alpha$'s and $\zeta_{FC}$'s.

\begin{figure}
	\centering
		\includegraphics[width=0.5\textwidth]{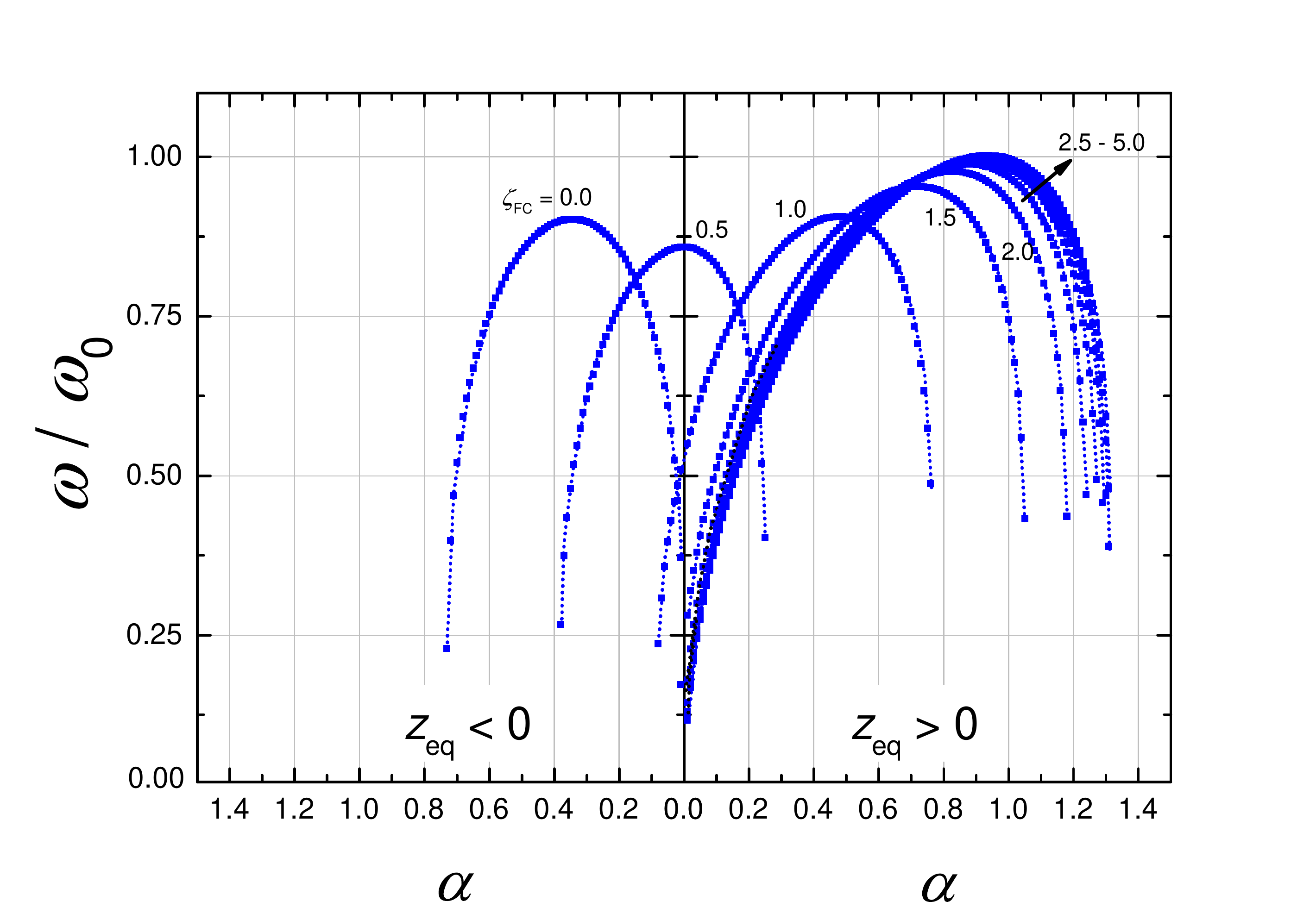}
	\caption{For a superconducting ring levitating in the field of a magnetic dipole, frequency of the trap around the equilibrium position as a function of the $\alpha$ parameter for different cooling positions (indicated in the plot). For simplicity, we show in the left (right) side the frequencies for the negative (positive) equilibrium positions. Note the decreasing scale in the left part ($\alpha$ is always positive).}
	\label{fig.freqfc}
\end{figure}

To study the stability of such solutions one can follow the same procedure as above. To simplify, we split the treatment in two parts: for the positive and for the negative equilibrium positions. In Fig. \ref{fig.freqfc} we present the results for the numerically evaluated frequency of the equilibrium position. For any $\zeta_{FC}$ there is an optimum $\alpha$ such that the equilibrium position is the most stable. We see also that the total maximum frequency is achieved in the ZFC case [$\omega_{opt}(\zeta_{FC}=\infty)\simeq1.004 \omega_0$], although for the other cooling positions, the optimum value of the frequency is above $0.85\omega_0$. In other words, the field cooling of the superconductor does not degrade substantially the stability of the equilibrium while it adds a new parameter to control the equilibrium position allowing a larger region for stabilizing the levitated ring. 

\section{Discussion}

\subsection{Lateral movements and rotational degrees of freedom}

One of the main objectives of the present work is to describe the forces resulting from the flux-conservation property of a levitated  superconductor. In this sense, we have focused on the main characteristics and properties that can be derived from this property and have concentrated on the \textit{vertical} movements and \textit{vertical} stability, this being the direction of the rings' axis.
Levitation is, however, only truly stable if it is so in all degrees of freedom, aside from rotation around the ring's axis. In a real ring with finite width and thickness, stability against rotation and lateral movements is provided by the current distribution induced on the ring's whole surface. This setting has recently been described and calculated numerically \cite{Latorre2020}.
Even though the finite thickness of the ring in any actual implementation may lead to stable levitation, it is still worthwhile studying the stability of the idealized case, and to explore further stabilizing mechanisms for levitation. This matter is discussed in the two following subsections.

\subsection{Stability in the idealized thin ring case}

For the (idealized) example of an infinitely thin ring, energy in a purely magnetic potential is minimized when the circulating current is zero. This is equivalent to preservation of the initial flux through the ring. In the ZFC case, the initial flux (zero) can be obtained at every position in space by tilting the ring until the local field is parallel to the surface.
The zero-flux condition can similarly be achieved in regions of the trapping field with an arbitrary field gradient or curvature: Rotation of a ring by 180 degrees about an axis in its plane will lead to a sign change of the flux in the ring's coordinate system. Since flux is a continuous quantity in free space, it follows immediately that an angle exists at which the flux must be zero.
For field cooling, the entire volume in which the average field is at least $\Phi_{FC}/\pi R^2$ is accessible, again by adjusting the orientation accordingly. The volume where the field is smaller than this value can not be reached without inducing a current.
This means that in either case, an idealized thin ring can not be trapped at the minimum of, for example, the field provided by anti-Helmholtz coils.

By extension, the above arguments are equally valid for an infinitely thin SC forming an arbitrary closed curve, no matter how intricate its shape.

One might conversely consider seeking trapping configurations which rely on a maximum of the flux, such that it decreases for any displacement or rotation. It follows from Earnshaw's theorem for the magnetic field that such a situation is impossible for static fields: Since there is no maximum of the magnetic field in free space, there can also be no maximum of the flux in a ring which is disconnected from all field sources.

\subsection{Stabilizing mechanisms}

Beyond finite-thickness effects, trapping of the levitator in the other degrees of freedom can be provided by other means, such as gyroscopic effects, time-orbiting potentials or more complex topologies.

Spin-stabilized magnetic levitation is well known from the physics of spinning devices like the Levitron \cite{Levitron}. We expect that similar considerations apply to a rotating ring.

A further path towards stable levitation of thin rings is the use of time-averaged potentials.
The combination of three non-orthogonal anti-Helmholtz coil pairs makes it possible to create a time averaged trapping potential, in which the flux through the ring is proportional to the distance from the trap center, regardless of its orientation. This configuration is equivalent to a ring rotating around both its planar axes in a static AHC field.
The rich dynamics of such quasi-static potentials presents fertile ground for future work.

In general, the idea of flux-conservation makes the levitation of a superconducting system 'rigid', in the sense that any variation of the flux in any element would make the superconductor react to counteract this variation.
This idea of rigid levitation has been exploited in several works of macroscopic levitation of superconductors, where the 'rigidity' comes from the hysteretic current penetration into the superconductor \cite{Brandt1990,Del-Valle2008,Del-Valle2009}. In a ring, it is the flux conservation which yields this 'rigidity', albeit only along one axis.
While a single ring cannot be trapped by the same mechanisms, the behaviour of a macroscopic superconductor can be approached by a rigidly connected arrangement of rings.  
A planar arrangement of electrically isolated loops will be pinned to its initial vertical and lateral position, albeit only in the FC case, since the zero-current condition can only be fulfilled there for all loops. Lowering the symmetry of the trap potential by applying different gradients in all directions can furthermore provide rotational stability. 

A minimal assembly of rings which allows stable levitation is found by using two non-parallel, rigidly connected rings. In the ZFC case, the flux through both rings can only be minimized when both are parallel to the local magnetic field, or both are in a plane in which the field components cancel over their areas. The first condition can be met, for example, by moving the rings along an axis until they reach the zero-current condition. As an example, if the rings are orthogonal to each other, their axes could lie along the lines $|x|=2z$ in an AHC field. Once again, applying a different field gradient along $y$ could provide rotational stability around the $z$-axis, however not against combinations of translation and rotation. These would be suppressed by the addition of gravity.
The use of multiple rings in a similar arrangement, e.g. placed on the three angled surfaces of a tetrahedron or the four angled surfaces of a pyramid, would be fully stable in an AHC-like field with different gradients along all axes.   
Complex, microscopic three-dimensional geometries made of superconducting materials pose a major challenge for fabrication, although recent work indicates that the creation of such structures is within reach of cutting-edge nano-assembly techniques \cite{Shani2020}.

\section{Conclusion}
\label{sec.conclusions}

We have introduced a theoretical framework for the levitation of superconducting rings over magnetic field sources, considering the current flowing in the superconductor resulting from the flux-conservation conditions (assumed as 1D circuit) as the main parameter. These conditions provide a novel set of properties with respect to those arising from the induced currents in a single-connected superconductor.

Our approach has been applied to both a quadrupolar anti-Helmholtz field, a case recently proposed for quantum magnetomechanics experiments, as well as to the case of a superconducting ring levitating in the field of a magnetic dipole. A complete set of analytical formulas are derived for describing these cases.
For the force acting on a ring in the $z$-direction, we have found that, depending on the cooling position, a continuous range of equilibrium positions can be achieved. We have also found the conditions that the system should satisfy in order to have this equilibrium; an optimum for having the largest possible vertical stability is obtained. This condition is expressed in terms of a single parameter that combines information from the levitating object (mass, radius, cross-section) as well as from the field source (magnetic moment). We have seen that even in the simple dipole-SC ring system, there is the possibility of having one, two, or zero vertical equilibrium positions.
The stability of the ring with respect to translation and rotation has been qualitatively discussed. Several strategies are proposed to achieve full stability. We have argued that although closed curved geometries are not trapped in DC fields, rigidly connected arrangements of several rings can be trapped, but only if they are three-dimensional or their symmetry is sufficiently low. In actual implementations, the finite thickness of the superconducting rings will lead to current distributions that result in stable levitation.

In quantum magnetomechanical experiments, the relevant dynamics of the system are those along a single axis. Our model takes the flux conservation through the ring into account and provides an analytical toolkit for the design and optimization of levitating rings.
The present work will guide future experiments of such levitation systems, 
by incorporating the degree of freedom provided by flux-conservation conditions as an extra key element for characterizing and enhancing the magnetic levitation.

\section{Acknowledgements}

We thank the members of the MaQSens Project for useful discussions.
We thank financial support from Catalan project 2014-SGR-150, Spanish project MAT2016-79426-P of Agencia Estatal de Investigaci\'on / Fondo Europeo de Desarrollo Regional (UE), the ESQ Discover Programme of the Austrian Academy of Sciences  (project ELISA), and the European Union Horizon 2020 Project FET-OPEN MaQSens (grant agreement 736943). AS acknowledges support from Catalan ICREA Academia program.

$^*$ Corresponding author: carles.navau@uab.cat.

\bibliographystyle{apsrev}
\bibliography{MuLevScenarios_v002,bibothers}

\end{document}